\documentclass[twocolumn,showpacs,preprintnumbers,amsmath,amssymb]{revtex4}

\usepackage{graphicx}
\usepackage{dcolumn}
\usepackage{bm}

\begin{document}


\title{The magnetic structure of Er$_2$Ti$_2$O$_7$}

\author{A.K.R. Briffa}
\author{R.J. Mason}
\author{M.W. Long}
\affiliation{%
School of Physics, Birmingham University, Edgbaston, Birmingham, B15 2TT, 
United Kingdom.
}%

\date{\today}

\begin{abstract}
We employ the previously published neutron scattering experiments\cite{1,2,3} 
to suggest that Er$_2$Ti$_2$O$_7$ has a broken-symmetry multiple-{\bf q} state 
with tetragonal magnetic symmetry\cite{4}.  The ordered moments do not appear 
to lie close to the crystal-field anticipated directions\cite{5} and we suggest 
that the low energy gapless mode, visible in specific-heat 
measurements\cite{6}, is not of the usual transverse isotropic Goldstone-mode 
type, but is longitudinal and is associated with the internal transfer of 
magnetism between distinct magnetic Bragg spots. 
\end{abstract}

\pacs{03.75.Lm, D75.25.+z, 75.30.Ds}
\maketitle

\section{\label{sec:level1}Introduction}

There is currently a lot of experimental and theoretical interest in rare-earth 
pyrochlore magnets of stoichiometry, R$_2$M$_2$O$_7$, where R is a rare-earth 
atom and M is an atom that has an available M$^{4+}$ state, such as tin or 
titanium.  Much of the interest has focused on `spin-ice'\cite{7}, the 
Ho$_2$Ti$_2$O$_7$ and Dy$_2$Ti$_2$O$_7$ compounds, due to the proposal that 
they can be considered to be analogue `monopole' systems\cite{8}.  The 
original motivation for studying these systems was the macroscopic degeneracy 
expected for the isotropic Heisenberg model\cite{9}.  The compounds 
Gd$_2$Ti$_2$O$_7$\cite{10} and Gd$_2$Sn$_2$O$_7$\cite{11} are systems that 
are expected to conform to this model at intermediate temperature.  The issue 
of how this residual degeneracy is lifted at very low temperature then 
surfaced and further work understanding the dipolar interaction was then 
relevant\cite{12}.  Spin-ice is a system with a strong crystal-field 
interaction forcing the spins to point along the natural crystallographic 
directions whereas the gadolinium compounds have essentially no anticipated 
crystal field (subject to some unexpected ESR issues\cite{13}).  To complete 
the set it would be nice to have a material where the local crystal-field 
promoted spins perpendicular to the natural crystallographic directions, and 
the compound Er$_2$Ti$_2$O$_7$ was thought to play this role\cite{14}.  
Theoretical crystal-field calculations\cite{5}, neutron scattering 
experiments\cite{1,2,3} and specific heat measurements\cite{6} were all 
brought forward to promote this picture.  Indeed, a subtle `order from 
disorder' calculation\cite{15} appeared to predict the observed experiments.  
In this article we offer an alternative view that the spins are oriented quite 
close to the natural crystallographic axes, but are actually slightly 
distorted away into a point-symmetry broken orientation.  This loss of 
point-symmetry then leads to a continuous degeneracy which explains the 
anomalous entropy still present at low temperatures.

Our proposal for the physics of Er$_2$Ti$_2$O$_7$ is essentially equivalent 
to previous studies of multiple-{\bf q} magnetism in face-centre-cubic 
magnets\cite{4}, which contain most of the fundamental ideas and concepts.
We organise the paper into three main sections;  firstly a discussion of the 
main experiments and how they lead us to our proposed ground-state, secondly 
a theoretical model that can be used to understand the majority of the 
experiments and thirdly a semi-classical spinwave analysis of our model.  Our 
model is clearly only approximately solved which causes clear `gaps' in our 
reasoning.  Finally we conclude.

\section{\label{sec:level2}Experiments and interpretation}

We commence with the specific heat measurements\cite{6}, which have been 
repeated as a function of field\cite{2,3,16}, although the original 
measurements are all that we require.  These specific heat measurements lead 
directly to three crucial physical facts.  Firstly there is a 
clear phase transition near in energy to the natural dipolar 
energy-scale\cite{14}.  Secondly there are only two states per atom involved in 
this low temperature phase transition, a Kramer's doublet, and so we need only 
deal with a pseudo-spin half.  Thirdly the relevant low temperature magnetic 
contribution is gapless and the observed power-law is consistent with a magnon 
dispersion that goes linearly to zero at isolated points in reciprocal-space.  
It is this third point which is totally unexpected:  spin-orbit interactions 
provide a coupling between the spin direction and the lattice which almost 
invariably lead to a spinwave gap.  In transition metals the spin-orbit 
interaction is weak in comparison to the exchange\cite{17} and so the gap is 
small, but in rare-earth magnets this interaction is usually huge and any 
associated low energy mode is absent.  The only likely failure for this 
argument is when the atom has a pure isotropic spin, gadolinium for example, where this 
anisotropy might be expected to be absent.  Erbium would be expected to have a 
huge anisotropy gap.  Sometimes low energy modes are seen (although not 
vanishing), in CeAs for example\cite{18}, however the interpretation for these 
modes is different\cite{4}.  These special modes occur in multiple-{\bf q} 
magnetism, only found in frustrated systems whose magnetism breaks the 
point-group symmetry of the underlying crystal.  The modes are associated with 
transferal of spin-density between distinct Bragg spots and correspond to local 
spin rotations which are longitudinal antiferromagnetic in character.  We are 
proposing that it is one of these special modes that has gone soft in 
Er$_2$Ti$_2$O$_7$.

The most important experiments for our development are those involving 
neutrons.  There is a powder diffraction experiment\cite{1}, a single-crystal 
diffraction experiment in a restricted plane but with critical inelastic 
measurements\cite{2}, a fairly complete single-crystal diffraction 
experiment\cite{3} and a polarised neutron study\cite{19}.  We will deal with 
the elastic scattering first and then consider the inelastic separately.  
Elementary investigation of neutron scattering data involves three critical 
elements:  the structure-factor, the form-factor and the orientational-factor.  
The structure-factor variation stems from the different moment directions 
occurring on distinct atoms in the magnetic unit cell, and for us is fairly 
definitive.  The form-factor arises from the orbital spread of the magnetic 
moment about the site of the nucleus, and is both a complication and a 
concern.  The magnetic erbium core is in a very complicated state, dominated by 
Hund's rules, it has a maximal total angular momentum which is parallel to its 
maximal spin to provide $J=15/2$.  The sixteen states are crystal field split 
to a relevant pseudo-spin doublet.  The additional crystal-field doublets are 
visible using inelastic neutron scattering\cite{1}, but are at high energies 
in comparison to the inate energy of the phase transition.  Indeed, the 
specific-heat measurements yield almost precisely this pseudo-spin degeneracy 
before the higher energy states impinge\cite{6}.  These 
complex low energy states have a shape that will distort the 
form-factor\cite{20}.  We will ignore this distortion and presume that the 
form-factor is spherically symmetric.  This issue is usually not a worry, 
because domain averaging reinstates the point-symmetry, however our proposed 
magnetism breaks that point-symmetry and would highlight these effects.  The 
orientational-factor (that stems from the fact that neutrons can only scatter 
from moments perpendicular to the direction of momentum transfer) is very 
physically informative and marks the start of our analysis.  

\begin{figure}
\includegraphics[height=7.2 cm, width=8.4 cm]{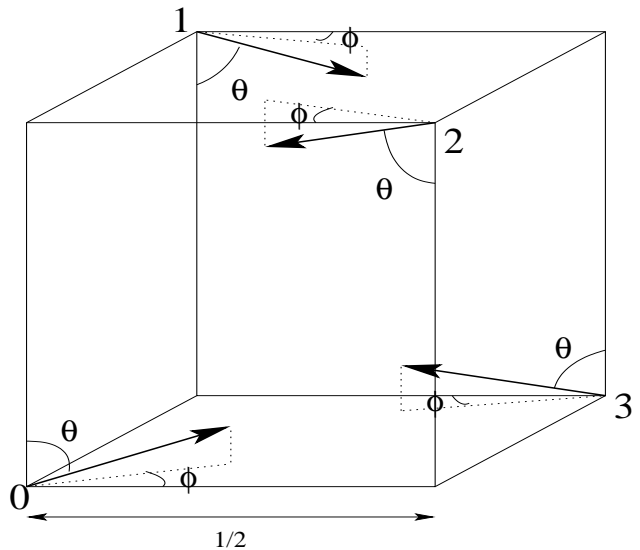}
\caption{\label{fig:1} The spin-states consistent with vanishing Bragg spots
(200), (020) and (002).}
\end{figure}

In simple terms, for any set 
of spots with the same structure-factor, one looks for the smallest 
(form-factor compensated) spot.  The magnetic moment associated with that 
structure-factor is then essentially parallel to the reciprocal-lattice-vector, 
{\bf G}, of that spot.  The neutron-scattering data then provides the 
following information.  The closest spots to the origin, $(111)$, $(1{\bar 1}{\bar 
1})$, $({\bar 1}1{\bar 1})$ $({\bar 1}{\bar 1}1)$, are all quite small.  The 
second-closest spots to the origin, (200), (020) and (002) are all very small, 
while the third-closest spots, (022), (202) and (220), are dominant.  The 
second observation is crucial, since it tells us that the three independent 
components of spin-density that characterise the magnetism are parallel to the 
Cartesian directions.  The states are described mathematically by
\begin{eqnarray}
\label{mag}
{\bf m}({\bf R})=\mid {\bf m}\mid \big[ e^{{\bf \hat x}.{\bf R}2\pi i}\sin 
\theta \cos \phi {\bf \hat x}\; \; \; \; \; \; \; \; \; \; \; \; &&\nonumber \\
\; \; \; \; \; \; \; \; \; \; \; \; +e^{{\bf \hat y}.{\bf R}2\pi i}\sin \theta 
\sin \phi {\bf \hat y}+e^{{\bf \hat z}.{\bf R}2\pi i}\cos \theta {\bf \hat z}
\big] &&
\end{eqnarray}
where ${\bf R}$ is the position of the atom in natural units (corresponding to 
the underlying face-centre-cubic Bravais lattice) and 
pictorially by Fig.\ref{fig:1}.  To limit to the depicted magnetism, we also 
require that there is no ferromagnetism in the system, which is controlled by 
the almost vanishing (222) Bragg spot.  Technically, these states are 
equivalent to Type I face-centre-cubic antiferromagnetism\cite{4}, but 
interestingly, the pyrochlore magnetism is easier to deduce than that of 
face-centre-cubic.  The key observation is that the $(111)$ spots provide 
complementary information:  this particular spot provides a phase which splits 
the lattice up into planes perpendicular to the $(111)$ direction, each of 
which has uniform phase.  These planes happen to be alternating Kagom\'e and 
sparse triangular planes with alternating signs.  The key observation is that 
these triangular planes make up one of the four sublattices, while the 
Kagom\'e planes make up the other three.  If the system is an antiferromagnet, 
then the structure factor for this Bragg spot should be proportional to the 
spin direction on the triangular sublattice.  The experimental observation 
that the $(111)$ Bragg spot is quite small then provides the information that 
the spin on site zero of Fig.\ref{fig:1} is almost parallel to the $(111)$ 
direction.  This clearly contradicts the endemic idea that the spins are 
perpendicular to the natural crystallographic 
directions\cite{1,2,3,5,14,15,16,19}.

The neutron experiments also clearly exhibit distinct magnetic 
domains\cite{1,2,3}.  This is clearly inconsistent with the idea that the 
state is a pure triple-{\bf q} state (spins pointing towards the natural 
crystallographic directions, defined by $\cos \Theta =\frac{1}{\surd 3}$), 
because this state preserves the cubic symmetry.  The magnetic field 
experiments provide us with an essentially unique state.  The experimental 
evidence was initially conflicting;  the first measurement\cite{1} tried to 
imply domains from a jump in intensity for a single Bragg spot when a field was 
applied.  Domains correspond to {\it redistribution} of Bragg intensity, and so 
if one peak grows then another should reduce, whereas a change in a single 
Bragg spot should signal a fundamental change in magnetism and {\it not} a 
change in domain structure.  This experimental result was therefore anomalous 
until the single-crystal measurement\cite{2} indicated an unexpected and 
unexplained diffuse peak that enlarged the Bragg spot with the inclusion of the 
field.  This single-crystal work\cite{2} does show clear indications of 
domains, but because not all of the magnetic Bragg scattering was observed it 
is not unambiguous.  The final fairly complete measurement\cite{3} measures all 
three crucial peaks independently and they are seen to separate in an intensity 
preserving way once the field is applied.  We can fit the three intensities to 
our Eq.(\ref{mag}) and we find that
\begin{equation}
\tan \phi =1\; \; \; \; \; \; \; \; \tan \theta \sim \frac{4}{3\surd 3}
\end{equation}
although no crystallographic significance should be attributed to our 
approximation.  This direction is quite close to $\frac{1}{\surd 6}(1,1,2)$, 
which also holds no special significance, unlike the previously proposed, 
$\frac{1}{\surd 6}(1,1,\bar 2)$, which is perpendicular to the natural 
crystallographic directions.  We have also analysed the $(111)$ spot\cite{2} 
which appears consistent with this choice, subject to our crude choice of 
form-factor, a pure hydrogenic $f$-state.  The state we propose has tetragonal 
magnetic symmetry and lies on the path that smoothly connects the pure 
triple-{\bf q} state to the single-{\bf q} state\cite{4}.

Finally, we consider the inelastic neutron scattering\cite{2} together with an 
ESR experiment\cite{16}.  The low-energy gapless mode is clearly visible close 
to the magnetic Bragg spots, and a second almost flat mode is also visible.  
Perhaps the most intriguing observation is that when the field is applied 
these magnons disappear, but reappear close to other point-group related Bragg 
spots\cite{2}, a subtle domain issue that we do not currently try to explain.  
Note that the position of this Goldstone mode is exactly equivalent to that 
in single-{\bf q} CeAs\cite{18} or indeed in USb\cite{21} or UO$_2$\cite{22}, 
both of which are thought to be pure triple-{\bf q} states.  In parallel to 
this, the ESR experiments\cite{16} see analogous modes but chase them as a 
function of field.  The physical surprise here is that it is the flat mode that 
softens and controls the loss of antiferromagnetism and not the low-energy 
Goldstone-mode as might na\"ively be presumed.  This is further evidence that 
the low energy mode is not of the usual type.

\section{Modelling}

Modelling rare-earth magnetism properly is complicated.  The dominance of 
Hund's rules makes the orbital nature of the states complex and the local 
crystal-field interactions destroy a simple isotropic-spin picture.  In 
previous work the case of Gd$_2$Ti$_2$O$_7$\cite{12} was considered, here we 
anticipate a pure spin with no orbital complications and natural models.  In 
this section we develop a phenomenological model to try to describe the 
observed behaviour of the erbium compound, and the natural models are severely 
renormalised.

One major theoretical complication is the crystal-field interaction, which 
projects the original large-J states onto a pseudo-spin half representation 
that controls the low temperature phase transition.  This projection strongly 
renormalises the interactions and hence corrupts intuition.  One can provide 
some understanding using an elementary model for this projection.  For a 
half-integer J we can easily show that the projection onto the states 
$\mid J,\pm J\rangle $ along the natural crystallographic directions, defined 
by,
\begin{eqnarray}
{\bf \hat z}_0\equiv \frac{{\bf \hat x}+{\bf \hat y}+{\bf \hat z}}{\surd 3}
\; \; \; \; \; \; \; \; \; \; 
{\bf \hat z}_1\equiv \frac{{\bf \hat x}-{\bf \hat y}-{\bf \hat z}}{\surd 3}
\; \; &&\nonumber \\
{\bf \hat z}_2\equiv \frac{-{\bf \hat x}+{\bf \hat y}-{\bf \hat z}}{\surd 3}
\; \; \; \; \; \; \; \; 
{\bf \hat z}_3\equiv \frac{-{\bf \hat x}-{\bf \hat y}+{\bf \hat z}}{\surd 3}&&
\end{eqnarray}
yields the mapping
\begin{equation}
\label{para}
{\bf \hat J}_\alpha \mapsto 2J{\bf \hat z}_\alpha \left( {\bf \hat z}_\alpha .
{\bf \hat S}_\alpha \right) 
\end{equation}
where ${\bf \hat J}_\alpha $ are the original angular-momentum operators and 
${\bf \hat S}_\alpha $ are the pseudo-spin half operators.  We can then 
project any bilinear Hamiltonian onto an effective spin-half model.  For the 
case of Er$_2$Ti$_2$O$_7$ the moment along the crystallographic axes is thought
to be small\cite{5} and in addition the ordered moment is indeed observed to be small\cite{1}.
To analyse this system it therefore seems more natural to
project onto the states $\mid J,\pm \frac{1}{2}\rangle $ using the following mapping:
\begin{equation}
\label{perp}
{\bf \hat J}_\alpha \mapsto \left( J+\frac{1}{2}\right) {\bf \hat S}_\alpha 
-\left( J-\frac{1}{2}\right) {\bf \hat z}_\alpha \left( {\bf \hat z}_\alpha .
{\bf \hat S}_\alpha \right) 
\end{equation}
The role of the crystallographic axes is dominant in the first case, 
Eq.(\ref{para}), and weakened in the second, Eq.(\ref{perp}). At 
bilinear order, our projection onto pseudo-spin is quite general, if the two 
coefficients in Eq.(\ref{perp}) are made parameters, depending only on the 
idea that the crystal-field is dominated by a single natural direction.

We develop five natural interactions in our modelling.  Four stem solely from 
the crystallographic directions and bilinear coupling.  The first, $H_0$, has 
no pseudo-spin analogue, and the others, $H_1$, $H_2$ and $H_3$, provide the 
dominant spin interactions.  The fifth interaction, $H_4$, couples the spin 
to the lattice and is crucial to our modelling.  We elect to derive these 
interactions by looking at the natural energies in our system.

The first energy we consider is the dipolar energy, this is not the strongest 
interaction as both exchange and residual crystal-field are expected to 
dominate, but it is very instructive.  Working on a single tetrahedron, in the 
presence of dominant antiferromagnetic Heisenberg interactions, the local 
dipolar interaction can be shown to take the form of a linear combination of 
two contributions
\begin{eqnarray}
H_4=\frac{1}{2}\Big[ \left( \hat S_0^x+\hat S_1^x-\hat S_2^x-\hat S_3^x\right) 
^2\; \; \; &&\nonumber \\+\left( \hat S_0^y-\hat S_1^y+\hat S_2^y-\hat S_3^y
\right) ^2\; \; \; &&\nonumber \\+\left( \hat S_0^z-\hat S_1^z-\hat S_2^z+\hat 
S_3^z\right) ^2\Big] &&
\end{eqnarray}
and
\begin{eqnarray}
H_0=\frac{1}{6}\Big[ \left( \hat S_0^x+\hat S_0^y+\hat S_0^z\right) ^2
+\left( \hat S_1^x-\hat S_1^y-\hat S_1^z\right) ^2\; \; \; \; \; \; \; \; &&
\nonumber \\ \; \; \; \; \; \; +\left( -\hat S_2^x+\hat S_2^y-\hat S_2^z\right) 
^2+\left( -\hat S_3^x-\hat S_3^y+\hat S_3^z\right) ^2\Big] &&\nonumber \\
\equiv \frac{1}{2}\sum _\alpha \left( {\bf \hat z}_\alpha .{\bf \hat S}_\alpha 
\right) ^2\; \; \; \; \; \; \; \; \; \; \; \; \; \; \; \; \; \; \; \; &&
\end{eqnarray}
where the coefficient of $H_0$ is six times that of $H_4$ for the local 
interaction.  Using Madelung techniques\cite{12}, it can also be shown that 
the dipole interaction, restricted to the ${\bf q}$=${\bf 0}$ subspace, also 
takes the same form (for the {\it static} issues) but with $H_0$ slightly 
more important.  We also elect to incorporate the interaction
\begin{equation}
H_3=\frac{1}{8}\left( \sum _\alpha {\bf \hat z}_\alpha .{\bf \hat S}_\alpha 
\right) ^2
\end{equation}
which is generated by our projections.  Treating these Hamiltonians as though 
they were initially functions of ${\bf \hat J}_\alpha $, applying the 
maximal spin projector then yields
\begin{equation}
H_4\mapsto 16J^2H_3\; \; \; \; \; H_0\mapsto 4J^2H_0\; \; \; \; \; H_3\mapsto 
4J^2H_3
\end{equation}
All that remains for this case is $H_0$ and $H_3$.  Although the Hamiltonian 
$H_0$ is crucial for the original angular momentum, once we arrive at our 
spin-half pseudo-spin, this interaction reduces to a constant and can be 
ignored.  As we only look at semi-classical solutions to our Hamiltonians, we 
are forced to keep $H_0$ active, although technically we need only the single 
Hamiltonian, $H_3$, for this limit.  Although $H_0$ is lost, there is still 
effectively a residual crystal-field interaction, $H_3$, promoting the natural 
crystallographic directions, it is just no longer local.  Spin-ice\cite{7} 
is thought to be well represented by the maximal projected pseudo-spin.  The 
spin component parallel to the crystallographic direction, ${\bf \hat z}_\alpha 
.{\bf \hat S}_j$, is then a conserved quantity and the pseudo-spin always 
points in these directions.  The residual interaction energy on a tetrahedron 
is described by $H_3$ and is minimised by a `two-in and two-out' 
configuration.  

We will now consider the minimal spin-projector model. It is slightly more 
complicated to analyse but provides
\begin{eqnarray}
\label{ten}
H_4\mapsto \left(J(J+1)+\frac{1}{4}\right) H_4-4\left( J(J+1)-\frac{3}{4}
\right) H_3&&\nonumber \\H_0\mapsto H_0\; \; \; \; \; \; \; \; \; \; \; \; \; 
H_3\mapsto H_3\; \; \; \; \; \; \; \; \; \; \; \; \; \; \; \; \; &&
\end{eqnarray}
The $H_3$ contribution now appears with the opposite sign. We observe that in the
large J limit Eq.(\ref{ten}) becomes
\begin{eqnarray}
H_4-4H_3=\; \; \; \; \; \; \; \; \; \; \; \; \; \; \; \; \; \; \; \; \; \; \; 
\; \; \; \; \; \; \; &&\nonumber \\
\frac{1}{12}\Big[ \left( \hat S_0^y-\hat S_1^y+\hat S_2^y-\hat S_3^y
-\hat S_0^z+\hat S_1^z+\hat S_2^z-\hat S_3^z\right) ^2\; \; \; &&\nonumber \\
+\left( \hat S_0^z-\hat S_1^z-\hat S_2^z+\hat S_3^z
-\hat S_0^x-\hat S_1^x+\hat S_2^x+\hat S_3^x\right) ^2\; \; \; &&\nonumber \\
+\left( \hat S_0^x+\hat S_1^x-\hat S_2^x-\hat S_3^x
-\hat S_0^y+\hat S_1^y-\hat S_2^y+\hat S_3^y\right) ^2\Big] &&
\end{eqnarray}
This provides a natural model which is minimised by
\begin{eqnarray}
\hat S_0^x+\hat S_1^x-\hat S_2^x-\hat S_3^x=\hat S_0^y-\hat S_1^y+\hat S_2^y
-\hat S_3^y&&\nonumber \\=\hat S_0^z-\hat S_1^z-\hat S_2^z+\hat S_3^z=A\; \; 
\; \; \; \; \; \; \; \; \; \; \; \; &&
\end{eqnarray}
There are many solutions to this, including the 
natural dipolar spirals (as found in Gd$_2$Sn$_2$O$_7$\cite{11}), the cubic 
symmetric triple-{\bf q} state and appropriate superpositions.

We now move on to exchange interactions.  Susceptibility experiments suggest 
that there is a fairly strong antiferromagnetic interaction\cite{23}, and for 
a single tetrahedron this is
\begin{eqnarray}
H_1=\frac{1}{2}\Big[ \left( \hat S_0^x+\hat S_1^x+\hat S_2^x+\hat S_3^x\right) 
^2\; \; \; &&\nonumber \\+\left( \hat S_0^y+\hat S_1^y+\hat S_2^y+\hat S_3^y
\right) ^2\; \; \; &&\nonumber \\+\left( \hat S_0^z+\hat S_1^z+\hat S_2^z+\hat 
S_3^z\right) ^2\Big] &&\nonumber \\\equiv \frac{1}{2}{\bf \hat T}.{\bf \hat T}
\; \; \; \; \; \; \; \; \; \; \; &&
\end{eqnarray}
defining a total-spin, ${\bf \hat T}$.  Although we have chosen to use a pure 
isotropic interaction, the true interaction probably depends on the details of 
the occupied orbitals and is consequently likely to be more complicated.  
Employing our maximal spin projection this reduces to
\begin{equation}
H_1\mapsto \frac{16}{3}J^2\left( H_0-H_3\right) 
\end{equation}
and, as expected, conflicts with `two-in and two-out'.  In a similar way to 
the previous interaction we can consider
\begin{eqnarray}
H_0-H_3=\frac{1}{24}\Big[ \left( \hat S_0^x+\hat S_0^y+\hat S_0^z-\hat S_1^x
+\hat S_1^y+\hat S_1^z\right) ^2\; \; \; &&\nonumber \\+\left( -\hat S_2^x+\hat 
S_2^y-\hat S_2^z+\hat S_3^x+\hat S_3^y-\hat S_3^z\right) ^2\; \; \; &&\nonumber 
\\+\left( \hat S_0^x+\hat S_0^y+\hat S_0^z+\hat S_2^x-\hat S_2^y+\hat S_2^z
\right) ^2\; \; \; &&\nonumber \\+\left( \hat S_1^x-\hat S_1^y-\hat S_1^z+\hat 
S_3^x+\hat S_3^y-\hat S_3^z\right) ^2\; \; \; &&\nonumber \\+\left( \hat S_0^x+
\hat S_0^y+\hat S_0^z+\hat S_3^x+\hat S_3^y-\hat S_3^z\right) ^2\; \; \; 
&&\nonumber \\+\left( \hat S_1^x-\hat S_1^y-\hat S_1^z+\hat S_2^x-\hat S_2^y+
\hat S_2^z\right) ^2\Big] &&\nonumber \\
=\frac{1}{16}\sum _{\alpha \beta }\left( {\bf \hat z}_\alpha .{\bf \hat 
S}_\alpha -{\bf \hat z}_\beta .{\bf \hat S}_\beta \right) ^2\; \; \; \; \; 
\; \; \; \; \; \; \; \; \; \; &&
\end{eqnarray}
which is the large-J limit.  This is minimised by
\begin{eqnarray}
\hat S_0^x+\hat S_0^y+\hat S_0^z=\hat S_1^x-\hat S_1^y-\hat S_1^z\; \; \; \; \; 
\; \; \; \; \; \; \; \; \; \; \; &&\nonumber \\=-\hat S_2^x+\hat S_2^y-
\hat S_2^z=-\hat S_3^x-\hat S_3^y+\hat S_3^z=B&&
\end{eqnarray}
providing another natural Hamiltonian.  Again there are many solutions to this, 
including all spins being both parallel and perpendicular to the natural 
crystallographic directions.  

Applying the minimal spin projection to $H_1$ provides
\begin{eqnarray}
H_1\mapsto \left( J+\frac{1}{2}\right) ^2H_1+\frac{4}{3}\left( J-\frac{1}{2}
\right) ^2\left( H_0-H_3\right) &&\nonumber \\-\left( J^2-\frac{1}{4}\right) 
H_2\; \; \; \; \; \; \; \; \; \; \; \; \; \; &&
\end{eqnarray}
where
\begin{equation}
H_2={\bf \hat T}.\sum _\alpha {\bf \hat z}_\alpha \left( {\bf \hat z}_\alpha .
{\bf \hat S}_\alpha \right) 
\end{equation}
Physically we see that the Heisenberg interaction remains dominant, but 
there are also terms equivalent to the maximal spin projector and additionally 
a new interaction, $H_2$, that would vanish for an antiferromagnet.  Note that 
the full dipolar interaction, restricted to a single tetrahedron, is 
proportional (up to a constant) to
\begin{equation}
H_4+6H_0-3H_2+\frac{7}{3}H_1
\end{equation}
and so all these interactions are `expected'.

Our dipolar arguments naturally lead to $H_0$, $H_1$, $H_2$, $H_4$ and (through 
pseudo-spin projection) $H_3$, but our use of the Heisenberg interaction is 
less natural.  There are likely to be uncontrolled orbitally driven distortions 
to the exchange process in this material.  We are consequently forced into 
treating our Hamiltonian as phenomenological.  Note that if for each exchange 
bond we simply enhance the bond when the spins are in the Cartesian plane of 
the bond, at the expense of when the spins are perpendicular to the plane, then 
we generate
\begin{eqnarray}
H=(1-\delta )\left[ \hat S_0^x\hat S_1^x+\hat S_2^x\hat S_3^x\right] \; \; \; 
\; \; \; \; \; \; \; \; \; \; \; \; \; \; \; \; \; \; \; \; \; \; \; 
&&\nonumber \\+(1+\delta )\left[ \hat S_0^y\hat S_1^y+\hat S_2^y\hat S_3^y+\hat 
S_0^z\hat S_1^z+\hat S_2^z\hat S_3^z\right] \; &&\nonumber \\+(1-\delta )\left[ 
\hat S_0^y\hat S_2^y+\hat S_1^y\hat S_3^y\right] \; \; \; \; \; \; \; \; 
\; \; \; \; \; \; \; \; \; \; \; \; \; \; \; \; \; \; &&\nonumber \\+(1+\delta 
)\left[ \hat S_0^z\hat S_2^z+\hat S_1^z\hat S_3^z+\hat S_0^x\hat S_2^x
+\hat S_1^x\hat S_3^x\right] &&\nonumber \\+(1-\delta )\left[ 
\hat S_0^z\hat S_3^z+\hat S_1^z\hat S_2^z\right] \; \; \; \; \; \; \; \; \; 
\; \; \; \; \; \; \; \; \; \; \; \; \; \; \; \; \; &&\nonumber \\+(1+\delta 
)\left[ \hat S_0^x\hat S_3^x+\hat S_1^x\hat S_2^x+\hat S_0^y\hat S_3^y
+\hat S_1^y\hat S_2^y\right] &&\nonumber \\=H_1-\delta H_4-2(1-\delta )S^2
\; \; \; \; \; \; \; \; \; \; &&
\end{eqnarray}
as desired in the next section.  The spin-orbit interaction can easily bias the 
exchange, which is mediated by orbital overlaps, in this way.

One experimental observation that is problematic for the state currently 
proposed in the literature is the severely reduced moment.  If the spins 
ordered in the proposed directions, perpendicular to the natural 
crystallographic axes, then one would expect to observe essentially the full 
moment, whereas a modest ordered moment fits the neutron Bragg data.  The 
previously proposed explanation for this small moment is quantum fluctuations, 
but our prediction has different physics.  We suggest that the spins align 
quite close to the natural crystallographic directions, where the pseudo-spin 
states have a modest observable moment, as suggested by Eq.(\ref{perp}), in 
agreement with the experiments.

\section{Spinwaves}

We elect to use the Hamiltonian
\begin{equation}
H=J\sum _t\left[ H_1-\delta H_4-\eta H_0+\xi (H_0-H_3)\right]
\end{equation}
where $t$ labels the tetrahedra, $J$ is the natural energy scale, $\delta $, 
$\eta $ and $\xi $ are scale-free parameters.  We choose to use {\it positive} 
values for these parameters (to stabilise the multiple-{\bf q} states) and 
consequently one should probably view this Hamiltonian as phenomenological.  
The residual crystal-field interaction can account for $\eta $ and $\xi $, but 
the choice of $\delta $ is physically opposite from the presumed dipolar 
source for this interaction.

This unexpected `failure' of the dipolar interaction has direct experimental 
consequences.  In real-space, point-size moments on a lattice have a dipolar 
energy proportional to
\begin{equation}
{\cal E}_D=-\frac{1}{2}\sum _{ii'}{\bf \hat S}_i.\frac{\partial }{\partial {\bf 
x}}{\bf \hat S}_{i'}.\frac{\partial }{\partial {\bf x}}V\left( \mid {\bf x}
\mid \right) \vert _{{\bf x}={\bf R}_i-{\bf R}_{i'}}
\end{equation}
where $V(X)$ is the Coulomb potential between all lattice sites but vanishes 
as $X\mapsto 0$.  In reciprocal space this transforms to
\begin{equation}
{\cal E}_D=\frac{1}{2}\sum _{\bf k}\sum _{\bf G}\mid \left( {\bf k}+{\bf G}
\right) .{\bf \hat S}_{\bf k}\mid ^2\tilde V(\mid {\bf k}+{\bf G}\mid )
\end{equation}
where $\tilde V(K)$ is the Fourier transform of $V(X)$.  Now since $V(X)$ is 
best represented as the Coulomb interaction multiplied by an increasing 
function, $\tilde V(K)$ corresponds to the Coulomb interaction multiplied by 
a reducing function.  To minimise the dipolar interaction we should minimise 
the contributions from the Bragg spots, ${\bf G}$, closest to the origin.  This 
means that the spin-density for these spots should be perpendicular to their 
reciprocal-space position.  Interestingly, because neutrons scatter with 
essentially an equivalent dipolar interaction, this means that these spots 
should be {\it maximal} losing nothing from the orientational factor in 
neutron scattering\cite{24}.  In Gd$_2$Sn$_2$O$_7$, and to a slightly lesser 
extent in Gd$_2$Ti$_2$O$_7$, these Bragg spots are indeed maximal\cite{11}, 
whereas in Er$_2$Ti$_2$O$_7$ they are either small or almost 
vanishing.  This tells us that the dipolar energy is intrinsically frustrated 
and explains why we are forced into using the apparently unphysical sign for 
$H_4$.

\begin{figure}
\includegraphics[height=7.2 cm, width=8.4 cm]{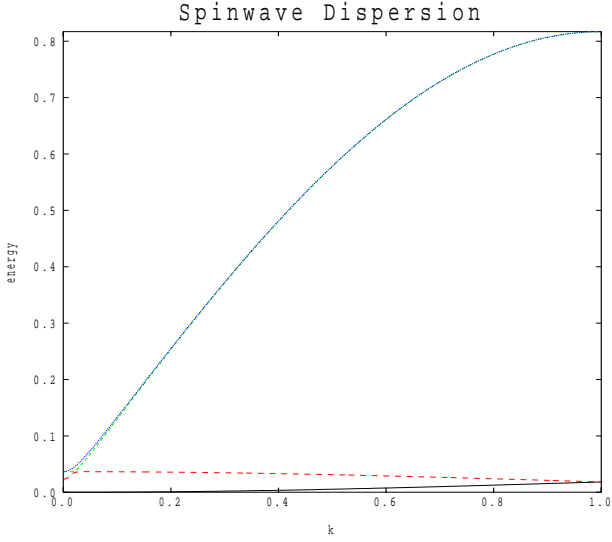}
\caption{\label{fig:2} The four spinwave branches for the case $\delta $=0.0001 
in units of $2JS$, parallel to one of the Cartesian axes for the case 
corresponding to the experimentally proposed ground-state.}
\end{figure}

\begin{figure}
\includegraphics[height=7.2 cm, width=8.4 cm]{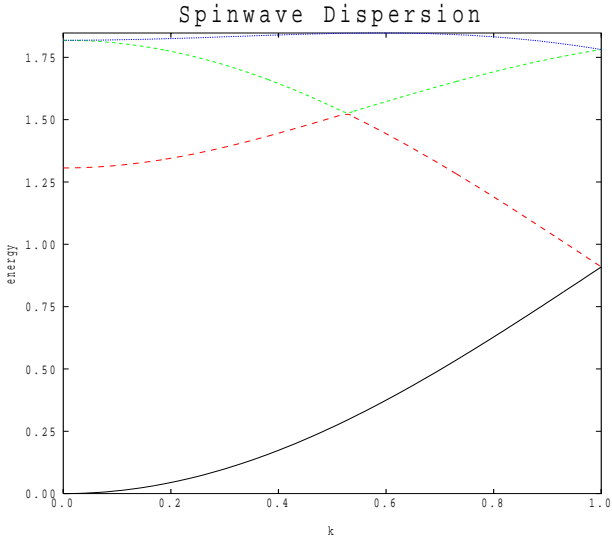}
\caption{\label{fig:3} The four spinwave branches for the case $\delta $=0.2 
in units of $2JS$, parallel to one of the Cartesian axes for the case 
corresponding to the experimentally proposed ground-state.}
\end{figure}

Since the system exhibits long-range order, we elect to solve our Hamiltonian 
in the semi-classical limit to assess the expected style of order and 
associated magnons.  We employ the Holstein-Primakoff transformation\cite{25}
\begin{equation}
\left[ \begin{matrix} \hat S^x\cr \hat S^y\cr \hat S^z\cr \end{matrix}\right] 
\mapsto \left[ \begin{matrix} \left[ \frac{S}{2}\right] ^\frac{1}{2}\left( 
b^\dagger +b\right) \cr \left[ \frac{S}{2}\right] ^\frac{1}{2}\left( b^\dagger 
-b\right) i\cr S-b^\dagger b\cr \end{matrix}\right] 
\end{equation}
with the $z$-axis oriented along the appropriate quantisation direction for 
each spin.  We start out solving the case $\eta $=0=$\xi $ to find that the 
classical solution of Fig.\ref{fig:1} denotes the ground-state manifold.  This 
then provides
\begin{equation}
{\bf \hat S}_0=\left[ \begin{matrix} {\bf \hat x}&{\bf \hat y}&{\bf \hat z}\cr 
\end{matrix}\right] \left[ \begin{matrix}\cos \theta \cos \phi &-\sin \phi 
&\sin \theta \cos \phi \cr \cos \theta \sin \phi &\cos \phi &\sin \theta \sin 
\phi \cr -\sin \theta &0&\cos \theta \cr \end{matrix} \right] \left[ 
\begin{matrix} \hat S_0^x\cr \hat S_0^y\cr \hat S_0^z\cr \end{matrix}\right] 
\end{equation}
\begin{equation}
{\bf \hat S}_1=\left[ \begin{matrix} {\bf \hat x}&{\bf \hat y}&{\bf \hat z}\cr 
\end{matrix}\right] \left[ \begin{matrix}\cos \theta \cos \phi &-\sin \phi 
&\sin \theta \cos \phi \cr -\cos \theta \sin \phi &-\cos \phi &-\sin \theta 
\sin \phi \cr \sin \theta &0&-\cos \theta \cr \end{matrix} \right] \left[ 
\begin{matrix} \hat S_1^x\cr \hat S_1^y\cr \hat S_1^z\cr \end{matrix}\right] 
\end{equation}
\begin{equation}
{\bf \hat S}_2=\left[ \begin{matrix} {\bf \hat x}&{\bf \hat y}&{\bf \hat z}\cr 
\end{matrix}\right] \left[ \begin{matrix}-\cos \theta \cos \phi &\sin \phi 
&-\sin \theta \cos \phi \cr \cos \theta \sin \phi &\cos \phi &\sin \theta \sin 
\phi \cr \sin \theta &0&-\cos \theta \cr \end{matrix} \right] \left[ 
\begin{matrix} \hat S_2^x\cr \hat S_2^y\cr \hat S_2^z\cr \end{matrix}\right] 
\end{equation}
\begin{equation}
{\bf \hat S}_3=\left[ \begin{matrix} {\bf \hat x}&{\bf \hat y}&{\bf \hat z}\cr 
\end{matrix}\right] \left[ \begin{matrix}-\cos \theta \cos \phi &\sin \phi 
&-\sin \theta \cos \phi \cr -\cos \theta \sin \phi &-\cos \phi &-\sin \theta 
\sin \phi \cr -\sin \theta &0&\cos \theta \cr \end{matrix} \right] \left[ 
\begin{matrix} \hat S_3^x\cr \hat S_3^y\cr \hat S_3^z\cr \end{matrix}\right] 
\end{equation}
for any particular tetrahedron.  Bloch transforming provides the excitations 
as
\begin{equation}
H=2JS\sum _{{\bf k}\in K_+}\left[ \begin{matrix}{\bf b}^\dagger _{\bf k}&
{\bf b}_{-{\bf k}}\cr \end{matrix}\right] \left[ \begin{matrix}A_{\bf k}&
B_{\bf k}\cr B^\dagger _{\bf k}&A_{\bf k}\cr \end{matrix}\right] \left[ 
\begin{matrix}{\bf b}_{\bf k}\cr {\bf b}^\dagger _{-{\bf k}}\cr \end{matrix}
\right] 
\end{equation}
where we are using the sublattice degrees of freedom as an implicit vector 
index and where $K_+$ denotes {\it half} of reciprocal-space, having divided 
out inversion.  The matrices are explicitly
\begin{widetext}
\begin{equation}
A_{\bf k}=\left[ \begin{matrix} 1+3\delta &-(\sin ^2\theta \cos ^2\phi 
+\delta )C_{y+z}&-(\sin ^2\theta \sin ^2\phi +\delta )C_{z+x}&-(\cos ^2\theta 
+\delta )C_{x+y}\cr -(\sin ^2\theta \cos ^2\phi +\delta )C_{y+z}&1+3\delta &
-(\cos ^2\theta +\delta )C_{x-y}&-(\sin ^2\theta \sin ^2\phi +\delta )C_{z-x}
\cr -(\sin ^2\theta \sin ^2\phi +\delta )C_{z+x}&-(\cos ^2\theta +\delta )
C_{x-y}&1+3\delta &-(\sin ^2\theta \cos ^2\theta +\delta )C_{y-z}\cr 
-(\cos ^2\theta +\delta )C_{x+y}&-(\sin ^2\theta \sin ^2\phi +\delta )C_{z-x}&
-(\sin ^2\theta \cos ^2\theta +\delta )C_{y-z}&1+3\delta \cr 
\end{matrix}\right] 
\end{equation}
\begin{equation}
B_{\bf k}=\left[ \begin{matrix} 0&(\cos \theta \cos \phi -i\sin \phi )^2C_{y+z}
&(\cos \theta \sin \phi +i\cos \phi )^2C_{z+x}&\sin ^2\theta C_{x+y}\cr 
(\cos \theta \cos \phi -i\sin \phi )^2C_{y+z}&0&\sin ^2\theta C_{x-y}&
(\cos \theta \sin \phi +i\cos \phi )^2C_{z-x}\cr (\cos \theta \sin \phi +i\cos 
\phi )^2C_{z+x}&\sin ^2\theta C_{x-y}&0&(\cos \theta \cos \phi -i\sin \phi )^2
C_{y-z}\cr \sin ^2\theta C_{x+y}&(\cos \theta \sin \phi +i\cos \phi )^2C_{z-x}&
(\cos \theta \cos \phi -i\sin \phi )^2C_{y-z}&0\cr \end{matrix}\right] 
\end{equation}
\end{widetext}
where
\begin{equation}
C_{\alpha \pm \beta }=\cos (k_\alpha \pm k_\beta )
\end{equation}
We can now diagonalise this Hamiltonian with a Bogoliubov transformation to 
construct the spinwave spectrum, $E_{{\bf k}\alpha }$.  In Fig.\ref{fig:2}
we look at the case where $\delta $ is very small.  There are two almost flat 
bands on the energy scale of $2JS\surd \delta $. These control the original 
degeneracy of the Heisenberg model which has been lifted by the inclusion of 
$H_4$.  In Fig.\ref{fig:3}
we take the case of a small but relevant $\delta $.
We see that the flat bands have now developed a little dispersion and 
are close to the highest energy excitations.  This plot has much in common 
with the experiments\cite{2}, but also has some clear problems.  The low energy 
mode is clearly visible, as is the almost flat high energy mode, but instead of 
a linear dispersion at low energy we find a quadratic dispersion.  This is 
because our model is actually degenerate across all the multiple-{\bf q} 
states of Fig.\ref{fig:1}, but if the particular ground-state were picked 
out by the energetics then the spinwave would be expected to harden.

\begin{figure}
\includegraphics[height=7.2 cm, width=8.4 cm]{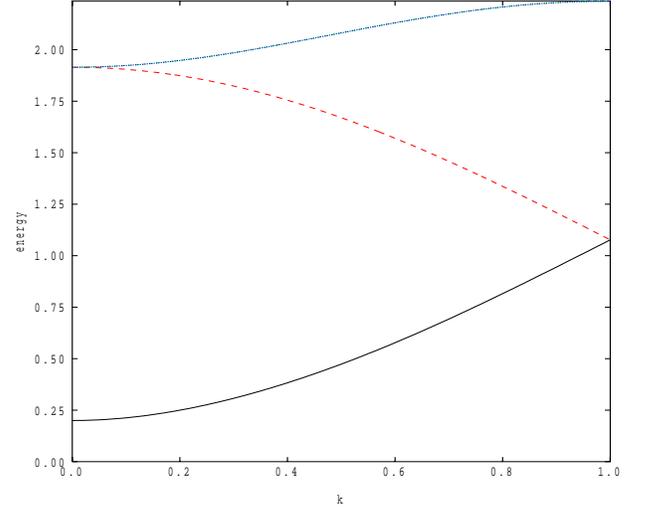}
\caption{\label{fig:4} The four spinwave branches for the case $\delta $=0.2, 
and $\eta $=0.2, which stabalises the cubic triple-{\bf q} state ($\phi 
=\frac{\pi }{4}$ and $\theta =\Theta $), in units of $2JS$, parallel to one of 
the Cartesian axes.}
\end{figure}

We are claiming that the low energy mode observed in the experiment is a rare 
form of longitudinal spinwave and so we now take time to explain this idea.  
We start out with an isotopic single-{\bf q} state, $\theta $=0, oriented along 
the $z$-axis.  The magnetic scattering would then appear at (220), but the 
(022) and (202) spots would be absent.  The usual Goldstone modes would be 
expected close to (002) and analogous spots on the body-centre-cubic 
superlattice generated by $\{ (2\bar 2\bar 2), (\bar 22\bar 2), (\bar 2\bar 22) 
\} $, but there would usually be a spinwave gap at (200) and (020), as these 
points are non-magnetic and unrelated to (002) by any magnetic symmetry.  The 
usual Goldstone modes are transverse; they are associated with the original 
state being tilted off axis and correspond to a small amount of magnetism 
at the same {\bf q}-point but in a perpendicular direction.  The initially 
gapped modes at (200) and (020) correspond to small magnetic distortions of 
the other two styles of point-group related magnetism which are currently not 
present in the system.  Physically they amount to small changes in the $\phi $ 
and $\theta $ of Fig.\ref{fig:1}.  The softening of these modes then 
corresponds to a {\it phase transition} from the original single-{\bf q} state 
into a multiple-{\bf q} state.  The incoming state naturally uses $\phi $ to 
become a double-{\bf q} magnet or $\theta $ to become a triple-{\bf q} 
magnet\cite{4}.  In a normal rare-earth material the spin-orbit interaction 
opens a huge gap in the standard transverse mode, but the longitudinal mode, 
which links to the other magnetic states can be found at low energies, as in 
CeAs\cite{18}.  The physical idea is then that as the angle $\theta $ smoothly 
rotates, from zero at the single-{\bf q} state to $\cos \Theta =\frac{1}{\surd 
3}$ in the cubic triple-{\bf q} state, the mode remains at zero energy and then 
generates a gap again in the cubic state.  Including the parameter $\eta $ into 
the spinwave calculation then stabilises the cubic triple-{\bf q} state  and 
provides the gapped spin-wave dispersion of Fig.\ref{fig:4}.

We now take time to offer an exactly solvable model which highlights some of 
the issues
\begin{equation}
H=-\delta H_4+\xi \left( H_0-H_3\right) 
\end{equation}
This model also has the multiple-{\bf q} states of Fig.\ref{fig:1} as its 
ground-state manifold, but it has the rather elementary dynamics controlled by
\begin{equation}
H=2JS\left[ \begin{matrix} \left[ \delta +\frac{\xi }{24}Z^*Z\right] C_{\bf k}&
\frac{\xi }{24}ZZC_{\bf k}\cr \frac{\xi }{24}Z^*Z^*C_{\bf k}&\left[ \delta 
+\frac{\xi }{24}Z^*Z\right] C_{\bf k}\cr \end{matrix}\right] 
\end{equation}
where
\begin{equation}
C_{\bf k}=\left[ \begin{matrix} 3&-C_{y+z}&-C_{z+x}&-C_{x+y}\cr -C_{y+z}&3&
-C_{x-y}&-C_{z-x}\cr -C_{z+x}&-C_{x-y}&3&-C_{y-z}\cr -C_{x+y}&-C_{z-x}&
-C_{y-z}&3\cr \end{matrix} \right] 
\end{equation}
and
\begin{equation}
Z=\cos \theta \sin \phi +i\cos \phi +\cos \theta \cos \phi -i\sin \phi -\sin 
\theta 
\end{equation}
The spinwave dispersion is given by
\begin{widetext}
\begin{equation}
E_{{\bf k}\alpha }=2JS\left[ \delta \left( \delta +\frac{\xi }{12}\left[ \left( 
\sin \theta \sin \phi -\cos \theta \right) ^2+\left( \cos \theta -\sin \theta 
\cos \phi \right) ^2+\left( \sin \theta \cos \phi -\sin \theta \sin \phi 
\right) ^2\right] \right) \right] ^\frac{1}{2}\epsilon _{{\bf k}\alpha }
\end{equation}
\end{widetext}
where
\begin{equation}
\epsilon _{{\bf k}\alpha }=4,4,2\pm \left[ 1+3\gamma _{\bf k}\right] 
^\frac{1}{2}
\end{equation}
and $\gamma _{\bf k}$ is the face-centre-cubic structure factor for the 
underlying periodicity.  The quantity $\epsilon _{{\bf k}\alpha }$ is 
essentially the pyrochlore lattice structure factor, exhibiting a gapless 
band and two flat bands describing the degeneracy (clear at finite energy).  
This model (ignoring the quadratic dependence at low energy) is enough 
to describe the observed inelastic neutron scattering\cite{2}, and hence this 
experiment does not clearly shed much light on {\it which} multiple-{\bf q} 
state might actually be stable.

Although all our multiple-{\bf q} states are classically degenerate, the 
quantum fluctuations lift this degeneracy\cite{26} and tend to stabilise 
collinear states.  For the current exactly solvable model we can calculate 
this quantum fluctuation energy analytically.  At zero temperature we find that 
the fluctuation energy per spin, $Q$, is
\begin{equation}
Q=3JS\left( -\delta -\frac{\xi }{24}\mid Z\mid ^2+\left[ \delta \left( \delta 
+\frac{\xi }{12}\mid Z\mid ^2\right) \right] ^\frac{1}{2}\right) 
\end{equation}
where
\begin{eqnarray}
\mid Z\mid ^2=3-\left( \sin \theta \cos \phi +\sin \theta \sin \phi +\cos 
\theta \right) ^2\; \; \; \; \; \; \; &&\nonumber \\=\left( \sin \theta \sin 
\phi -\cos \theta \right) ^2+\left( \cos \theta -\sin \theta \cos \phi \right) 
^2&&\nonumber \\+\left( \sin \theta \cos \phi -\sin \theta \sin \phi \right) ^2
\; \; \; \; \; \; \; \; \; \; \; \; \; \; \; \; \; &&
\end{eqnarray}
This fluctuation energy vanishes for the cubic triple-{\bf q} state and is minimised when the 
spins are perpendicular to the natural crystallographic directions.  This 
exactly solvable model has very anisotropic interactions which do not promote 
standard collinear states.  If we accept that our spin is pseudo-spin half 
and that the quantum fluctuation energy is of the same order as the classical 
energy, then we can now attempt a physical explanation for the system.  If 
there is a classical energy promoting the cubic triple-{\bf q} state, such as 
$H_0$ or $H_3$, then this could balance the quantum fluctuation energy at the 
observed experimental state.  However, careful analysis of the current model 
shows that the state jumps discontinuously from the cubic state to one with the 
spins perpendicular to the crystallographic directions and so the current model 
is too simplistic.

\begin{figure}
\includegraphics[height=7.2 cm, width=8.4 cm]{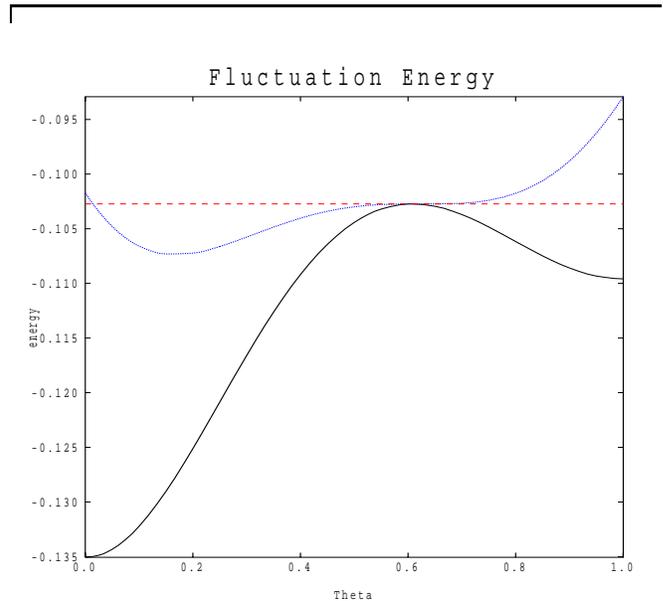}
\caption{\label{fig:5} The quantum fluctuation energy for the case $\delta 
$=0.2, as a function of $\theta $ (in units of $\frac{\pi }{2}$)with 
$\phi $=$\frac{\pi }{4}$ (solid curve). We have included a small $\eta $ to 
achieve the blue curve.}
\end{figure}

Returning to our more physical model, $H_1-\delta H_4$, we can calculate the 
quantum fluctuation energy as
\begin{equation}
Q=JS\left( -1-3\delta +\frac{1}{4N}\sum _{{\bf k}\alpha }E_{{\bf k}\alpha }
\right) 
\end{equation}
at zero temperature, and we offer an example in Fig.\ref{fig:5}.
As expected, the ground-state is now the single-{\bf q} state (black curve).  We next add 
a carefully scaled classical contribution favouring the cubic state and now an 
intermediate state appears relatively stable (blue curve).  However the energetics are not very 
`stable' and rather than continuing to evolve smoothly, the ground-state tends to make a small 
discontinuous jump when close to the cubic state, to the cubic state.

Having analysed a particular case, we now move on to generalities.  Our 
complete Hamiltonian is
\begin{eqnarray}
H=J\sum _t\Big[ \alpha H_1-\delta H_4-\eta H_0+\xi (H_0-H_3)&&\nonumber \\
-\frac{3}{2}\beta H_2+\epsilon H_5\Big] \; \; \; \; \; \; \; \; &&
\end{eqnarray}
where $H_5$ is a local interaction which lifts the azimuthal symmetry and 
provides a crystal-field preference appropriate to the weak hexagonal symmetry 
around the natural crystallographic directions.  We can semi-classically 
solve such an interaction in general.  There are four natural ground-states:  
the dipolar spirals\cite{11}, a generalised `two-in and two-out' state (typical 
for spin-ice\cite{7}), the pure triple-{\bf q} state\cite{4} and finally a 
multiple-{\bf q} state with the spins perpendicular to the natural 
crystallographic directions, but pointing along one of the twelve natural 
directions allowed by the weak hexagonal interaction.  The key observation is 
that in {\it all} of these phases there is a natural gap in the spinwave 
spectrum.  Our generic Hamiltonian is very anisotropic and that generates 
the expected loss of continuous symmetry and a consequent gap.  The observed 
gapless state is truly unexpected.

There is no easy way to eliminate the gap in the spiral and `two-in and 
two-out' phases, but since Er$_2$Ti$_2$O$_7$ has a multiple-{\bf q} 
ground-state this is irrelevant.  One way to close the gap is to omit the 
hexagonal crystal-field interaction, $\epsilon \mapsto 0$. This is the current 
core of the picture prevalent in the literature\cite{1,2,3,15}.  If it were 
exact, then one could use a magnetic field to access this degeneracy but the 
current experiments point to the particular azimuthal angle being pinned and 
nonreactive to a field.  A weak hexagonal field provides a small gap. However it is 
possible that the observed linear dispersion, and specific heat, have not been accessed at low 
enough energies, and temperatures respectively, to observe this gap.
Another way to close the gap, our explanation, is that the 
multiple-{\bf q} degeneracy is active.  We would leave the hexagonal 
interaction, which stabilises $\phi =\frac{\pi }{4}$, but omit the excess 
local crystal-field term, $\eta \mapsto 0$.  The inclusion of the hexagonal 
interaction removes one inconsistency in our model; the quadratic dispersion at 
low energy becomes linear as required, and our continuous degeneracy is in the 
angle $\theta $ and not $\phi $.  We believe that once the particular choice 
of $\theta $ is chosen by some collective compromise, the gaplessness still 
remains.  Since there is no such collective competition in our semi-classical 
analysis, we have resorted to suggesting that quantum fluctuations could 
provide such an asymmetric $\theta $, without offering any rigorous 
calculations to prove this.

\section{Conclusions}

Rare-earth magnets are usually very anisotropic with a gap to magnetic 
excitations.  Er$_2$Ti$_2$O$_7$ is an unusual example of a gapless rare-earth 
magnet.  The experimental evidence points towards a multiple-{\bf q} 
antiferromagnet where the spins do not point along crystallographically natural 
directions.  The low energy mode then smoothly varies this incommensurate 
angle.  The single-domain single-crystal measurements indicate that the angles 
$\phi $=$\frac{\pi }{4}$ and (the incommensurate) $\theta \sim 
\frac{\pi }{5}$ are selected.

The observed spinwave spectrum is well represented by the natural pyrochlore 
structure factor and hence does not restrict the model much, except that the 
spectrum is gapless and so a continuous degeneracy is required.  The observed 
orientational factors in the neutron scattering are contradictory to the 
expectations of the dipolar interaction and this leads us to propose the model 
with the opposite sign for $H_{4}$ in order to fit the experiments.

Our physical insight arises from the balance between quantum fluctuations and 
classical interactions.  Our Hamiltonian, $H_1$-$\delta H_4$, is classically 
degenerate, with the states of Fig.\ref{fig:1} as the ground-state manifold.  
The quantum fluctuations stabilise the single-{\bf q} state, and because the 
pyrochlore lattice is highly frustrated the quantum fluctuation energy is 
expected to be large.  Due to the strong spin-orbit and crystal-field effects 
there are additional interactions which happen to favour the natural 
crystallographic directions.  If we imagine increasing such an interaction 
smoothly from zero, at some strength it will overturn the quantum fluctuation 
energy and drive the system into a multiple-{\bf q} state.  We are proposing 
that this transition would be second-order and would pass through the 
intermediate gapless phases where $\theta $ varies from zero to $\Theta $, with 
$Er_{2}Ti_{2}O_{7}$ part-way through this process at the observed 
value of $\theta \sim \frac{\pi }{5}$.

Our proposal requires a multiple-{\bf q} state with an unnatural angle 
$\theta $ which is not clearly derived nor convincingly explained.  The current 
picture in the literature has the natural angles $\theta $=$\frac{\pi }{2}$ and 
$\phi $=$\frac{\pi }{4}$ and so would be expected to have a gap.  
This gap needs to be too small to be observed.  This state should also have 
a dominant (111) spot, as this Bragg spot structure factor is perpendicular 
to its momentum transfer, whereas the spot is observed to be small.  The only 
weakness in our argument is the form-factor, which would have to have a severe 
and unexpected orientational dependence.

There are now two plausible states that will fit the experiments:  the original 
idea\cite{15}, with spins perpendicular to the natural crystallographic 
directions, and our proposal with a non-symmetric angle $\theta $.  How can 
we decide between them experimentally?  The original proposal should exhibit 
a gap, controlled by the hexagonal crystal field, which can be sought.  If 
this gap is tiny, then an external magnetic field should be able to rotate 
the angle $\phi $ and dramatically change the magnetic Bragg scattering, with 
the ratio of intensities changing from $(1,1,4)$ to $(3,3,0)$ for the three 
main Bragg spots.  One would also need an explanation for why the $(111)$-type 
Bragg spots were so small, either from dramatic form-factor dependence or from 
some other as yet unresolved experimental issue.  Our proposal has rather 
different issues.  Since we fit the known data the experiments are not too 
pertinent, it is the theory that is critical.  We need a model that stabilises 
the unsymmetric state and we need the state to remain gapless, both issues 
having been side-stepped in this article.

Our proposed model is a pseudo-spin half model on the pyrochlore lattice but 
our semi-classical magnetic techniques are not sophisticated enough to predict 
our expectations.  Better theoretical analysis of the quantum mechanical model 
is required to make a more detailed comparison with experiment.

\begin{acknowledgments}
We wish to acknowledge useful discussions with E.A. Blackburn.
\end{acknowledgments}


\end{document}